\documentclass[aps,prb,twocolumn,floatfix,superscriptaddress]{revtex4-2}
\usepackage[utf8]{inputenc}
\usepackage{mathrsfs}
\usepackage{graphicx}
\usepackage[english]{babel}
\usepackage{amsmath}
\usepackage{amssymb}
\usepackage{xcolor}
\usepackage{relsize}
\usepackage{CJK}
\usepackage{textcomp}
\usepackage{dsfont}
\usepackage{bm}

\newcommand{\me}{\mathrm{e}}
\newcommand{\mi}{\mathrm{i}}

\newcommand{\dif}{\mathrm{d}}

\allowdisplaybreaks
\begin{document}

\title{Bound-like State in a 1D Self-Similar Delta-Barrier Array}

\author{Jia-Chen Tang}
\affiliation{School of Physics, Southeast University, Jiulonghu Campus, Nanjing 211189, China}
\author{Xu-Yang Hou}
\affiliation{School of Physics, Southeast University, Jiulonghu Campus, Nanjing 211189, China}
\author{Yan He}
\affiliation{College of physics, Sichuan University, Chengdu, Sichuan 610064, China}

\author{Hao Guo}
\email{guohao.ph@seu.edu.cn}
\affiliation{School of Physics, Southeast University, Jiulonghu Campus, Nanjing 211189, China}
\affiliation{Hefei National Laboratory, Hefei 230088, China}
\begin{abstract}
We investigate a one-dimensional quantum system with a self-similar arrangement of delta-function potential barriers, exhibiting discrete scale invariance. The singular potential induces kinematically enforced symmetry breaking at $x=0$, decoupling the positive and negative spatial regions and leading to non-symmetric zero-energy states. We demonstrate that the system supports a unique zero-energy wavefunction, which, though not square-integrable, decays to zero at infinity and acts as a bound-like state with self-similar properties under discrete scaling transformations, akin to Efimov physics but limited to a single state. In momentum space, this wavefunction exhibits a threshold singularity at low momenta, with behavior depending on the scaling exponent $\alpha$:power-law divergence and log-periodic modulations for $0 < \alpha < 1$, logarithmic divergence for $\alpha = 1$, and a finite limit for $\alpha > 1$, which may be observable through time-of-flight or spectroscopic measurements in cold atom experiments. The system's continuous spectrum, starting at zero energy, lacks discrete bound states. These findings highlight the role of singular potentials in generating scale-invariant quantum phenomena and provide a minimal framework for studying discrete scale symmetry and its potential experimental signatures.
\end{abstract}

\maketitle

\section{Introduction}

Quantum systems exhibiting scale invariance are of fundamental interest due to their unique mathematical structure and nontrivial physical implications, particularly in the statistical physics and condensed matter physics\citep{PhysRevA.109.043301, PhysRevE.103.012117,annurev102419-032845,PhysRevA.86.012710,10.1119/5.0086802,PhysRevLett.121.023401,PhysRevA.100.011403,Hammer2008,Nishida2013,Naidon_2017}. In such systems, their physical laws remain invariant under coordinate rescaling, which often leads to distinctive features such as wavefunctions that retain structural similarity under scaling transformations and energy spectra that follow power-law or even geometrically spaced distributions. A paradigmatic example is the inverse-square potential $V(x) = V_0/x^2$, which exhibits continuous scale invariance and serves as a textbook model for scale-invariant quantum mechanics\cite{case1950}.

In quantum few-body physics, the manifestation of discrete scale invariance (DSI) is most famously embodied in the Efimov effect\citep{efimov1970energy}, first predicted in 1970. Efimov demonstrated that three particles with resonant interactions can form an infinite series of self-similar bound states whose energies are related by a geometric scaling factor. This effect was long considered a purely theoretical curiosity until its experimental verification in ultracold atoms, which sparked renewed interest in the broader implications of DSI in quantum systems\citep{BRAATEN2006259,PhysRevLett.103.130404,kraemer2006evidence, PhysRevLett.102.165302, Pollack2009, PhysRevLett.107.120401, lompe2010radio, PhysRevLett.112.250404,  PhysRevLett.111.053202, zaccanti2009observation,PhysRevLett.112.190401}. Recently, the concept of the Efimov effect has been extended to many-body dynamics and condensed matter systems, such as the ``Efimovian expansion'' observed in scale-invariant Fermi gases \citep{deng2016observation} , and bound-like states induced by electron-charged impurity interactions in Dirac semi-metals \citep{zhang2018efimov}. Furthermore, the discrete scaling symmetry in quantum systems with a $1/r^2$ potential can lead to dynamical fractals in the time domain, manifesting as self-similar patterns in observables like the Loschmidt amplitude \cite{PhysRevLett.122.230402}. These studies suggest that the Efimov effect is not confined to few-body systems but may manifest as dynamic or bound-like states in a broader range of quantum systems.

Although significant progress has been made in the study of scale-invariant systems, most models still rely on smooth and continuous potentials, such as the Coulomb potential. In contrast, the role of singular potential structures in realizing scale invariance has not been thoroughly explored.
Therefore in this work, we explore an alternative route to scale invariance by constructing a quantum system with a series of delta-function barriers arranged in a self-similar manner. This singular potential structure introduces discrete scale invariance, leading to interesting physical behaviors not observed in conventional scale-invariant systems.
The model we consider is a one-dimensional quantum system with a Hamiltonian of the form $ H = -\frac{\hbar^2}{2m} \frac{\dif^2}{\dif x^2} + V(x) $, where the potential $ V(x) $ consists of delta-function barriers placed at positions $ x_n = x_0 \lambda^n $ for $ n \in \mathbb{Z} $, with $ \lambda > 1 $ being the scaling factor. This arrangement ensures that under a spatial rescaling $ x \to \lambda x $, the potential transforms in a way that preserves the form of the Hamiltonian up to a multiplicative factor, thereby endowing the system with discrete scale invariance.

A key feature of this potential is the accumulation of infinitely many delta-function barriers near the origin, creating an effective infinite barrier at $ x = 0 $. This results in the decoupling of the positive and negative halves of the real line, allowing us to focus our analysis on the region $ x > 0 $. We investigate the zero-energy wavefunction of this system, which, despite not being square-integrable, exhibits intriguing properties such as self-similarity and power-law decay.
In momentum space, this wavefunction displays a power-law divergence at $0<\alpha<1$ with log-periodic modulations, offering experimental signatures.
These characteristics make it a bound-like state, providing a bridge between bound and scattering states in quantum mechanics.

Our study not only highlights the role of singular potentials in generating scale-invariant behavior but also underscores the importance of discrete symmetries in quantum systems. By analyzing the recursive properties of the wavefunction across different intervals defined by the potential barriers, we uncover the conditions under which the system supports such bound-like states. Furthermore, we discuss the implications of these findings for the spectral properties of the Hamiltonian, particularly the absence of a discrete spectrum and the presence of a continuous spectrum starting from zero energy.

This work may broaden the understanding of scale invariance in quantum mechanics. The results presented here can find applications in various areas of physics where self-similar structures and singular potentials play a crucial role, such as in the study of fractals, disordered systems, and quantum field theory.

The rest of this paper is organized as follows. In Section~\ref{Sec2}, we present the model of a one-dimensional quantum system featuring self-similar delta-function barriers, examining its scale symmetry and spectral characteristics, while highlighting the spatial decoupling of the wavefunction induced by an infinitely high barrier at $x=0$. Section~\ref{Sec3} focuses on the zero-energy wavefunction, analyzing its structure through a recursive transformer matrix approach and exploring its scale invariance, asymptotic behavior, normalization challenges, and the momentum-space behavior, including power-law divergence and log-periodic modulations at $0<\alpha<1$ which is observable in experiments. Finally, Section~\ref{Sec4} summarizes our findings and offers concluding remarks.
Moreover, we will adopt natural units, setting $\hbar =  1$, to simplify calculations and notation.

\section{1D System with a self-similar potential well}\label{Sec2}
\subsection{Mirror symmetry}
We consider the simplest one-dimensional scenario, where the Hamiltonian is given by $H(x)=-\frac{1}{2m}\frac{\dif^2}{\dif x^2}+V(x)$. Under the scaling transformation $x\to \lambda x$, the kinematic term transforms as $-\frac{1}{2m}\frac{\dif^2}{\dif x^2}\to \frac{1}{\lambda^2}\left(-\frac{1}{2m}\frac{\dif^2}{\dif x^2}\right)$, reflecting its characteristic length dimension of $[L]^{-2}$. To maintain scale invariance of the entire Hamiltonian, the potential term must exhibit identical scaling behavior. This requirement uniquely selects the inverse-square potential: $V(x)=\frac{V_0}{x^2}$ if it is a well-behaved function. However, by allowing discontinuous potentials, an alternative construction becomes possible.  Noting that the Dirac $\delta$-function $\delta(x)$ carries length dimension $[L]^{-1}$, we can construct a scale-invariant system through a carefully designed singular potential with discrete scale symmetry:
\begin{equation}\label{V1}
V(x) = V_0 \sum_{n=-\infty}^{\infty} \delta(x^2 - x_n^2) , \quad x_n = x_0 \lambda^n,
\end{equation}
where $ V_0 $ represents the strength of each delta potential well, $ \lambda > 1 $ is the scaling factor, and $ x_0 >0$ denotes the initial distance unit.

Due to the presence of infinitely dense $\delta$-function potential barriers near $x = 0$, with their strengths diverging toward infinity, the system effectively forms an infinitely high barrier at $x = 0$, rendering this point impenetrable to particles. Consequently, $x = 0$ acts as a hard wall, separating space into two independent semi-infinite regions: $x > 0$ and $x < 0$. This decoupling holds at all energy levels, including the ground state (if it exists). Therefore, the ground-state wavefunctions in the regions $x > 0$ and $x < 0$ can be treated as two independent solutions, since quantum tunneling across the barrier is entirely suppressed. This phenomenon resembles spontaneous symmetry breaking (SSB) in that the Hamiltonian retains mirror symmetry, but the physical states do not. However, in contrast to conventional SSB, which typically involves degenerate ground states and dynamical mechanisms such as interactions or energy minimization, the symmetry breaking in our system is driven solely by the geometric and boundary constraints imposed by the singular potential. Crucially, this symmetry breaking does not arise from the system's internal dynamics or a spontaneous selection among degenerate states; rather, it is directly enforced by the specific geometric arrangement of the potential and the boundary conditions at $x=0$. The wavefunction cannot be symmetrically extended across the origin due to its non-normalizability, which will be elaborated upon later, and the breakdown of continuity at $x=0$.

We therefore designate this phenomenon as the kinematically enforced symmetry breaking. It is analogous to symmetry breaking via self-adjoint extension choices in singular quantum mechanical problems \cite{simon1976}, and similar to how strong boundary conditions in double-well or inverse-square potentials can eliminate symmetric states from the physical Hilbert space \cite{gridnev2016,callan1977}.
Such symmetry breaking mechanisms, while not common in textbook quantum mechanics, do appear in systems with singular potentials, impenetrable barriers, and effective decoupling of spatial regions. Notably, similar symmetry reduction arises in Efimov physics through boundary condition selections at short distances \cite{nishida2012}, and in edge-state localization in topological chains \cite{ryu2002}. The physical implications of this symmetry breaking will be discussed in subsequent sections.

\subsection{Scaling symmetry}

Note that the potential can be recast in the form
\begin{align}\label{V2}
V(x) = V_0 \sum_{n=-\infty}^{\infty} \frac{\delta(x - x_n) + \delta(x + x_n)}{2|x_n|},
\end{align}
where the coefficient $\frac{V_0}{2|x_n|} = \frac{V_0}{2x_0\lambda^n}$ represents the effective coupling strength associated with the point-like barriers located symmetrically at $x = \pm x_n$. This construction differs fundamentally from the well-known Kronig-Penney model \cite{KronigPenney1931}, where the barriers are spaced periodically. In contrast, the present model features an exponentially increasing separation between adjacent barriers as one moves away from the origin.
The resulting potential describes an infinite array of $\delta$-function barriers with spatially modulated strength. Under the rescaling transformation $x \to \lambda x$, the set of barrier positions transforms as $x_n \to x_{n+1}$, thereby endowing the system with discrete scale invariance.
In fact, the Hamiltonian satisfies the scaling relation $H(\lambda x) = \lambda^{-2} H(x)$, as elaborated in Appendix~\ref{app1}. This self-similar structure, characterized by the dimensionless scaling factor $\lambda$, eliminates an absolute length scale from the physical observables of the system, giving rise to a range of unconventional features in the behavior of wavefunctions. For instance, if $\psi(x)$ is an eigenstate of $H(x)$ with energy $E$, then it appears that $\psi(x/\lambda)$ is also an eigenstate, but with energy $E/\lambda^2$ if it is well-behaved (see Appendix~\ref{app1}). This scaling can be iterated indefinitely. For any positive integer $n$, define $\psi_{\lambda^n}(x) := \psi\left(x / \lambda^n\right)$, which satisfies \begin{align} H(x) \psi_{\lambda^n}(x) = \frac{E}{\lambda^{2n}} \psi_{\lambda^n}(x). \end{align}
If an energy $E$ exists in the spectrum, all scaled energies $\lambda^{-2n} E$ (where $n \in \mathbb{Z}$) are also included, resulting in a dense spectrum for $E \geq 0$. However, $\psi_{\lambda^n}(x) := \psi(x / \lambda^n)$ is not necessarily a valid eigenstate, despite the Hamiltonian's formal scale invariance. Eigenstates must satisfy boundary conditions, normalizability, and matching conditions at the $\delta$-function potentials, which are well-defined (see Appendix~\ref{app1}). Thus, the set of eigenstates is not guaranteed to be closed under scaling transformations. We will return to this subtlety in a later discussion.

The resulting potential describes an infinite array of $\delta$-function barriers with spatially modulated strength, endowing the system with discrete scale invariance characterized by the dimensionless scaling factor $\lambda$, which eliminates an absolute length scale in the physical observables, such as the wavefunction's scaling exponent and spectral properties.

\subsection{The spectrum}

As $ n \to -\infty $, the positions $ x_n \to 0 $, leading to an accumulation of infinitely dense $\delta$-function barriers near $ x = 0 $, each with strength $ \frac{V_0}{|x_n|} $. If $ V_0 < 0 $, then the barrier strength diverges to negative infinity, resulting in a potential that is unbounded from below and singular at the origin. In such cases, the system typically lacks a ground state. For this reason, our analysis will primarily focus on the case where $ V_0 > 0 $.

For $ V_0 > 0 $, it can be shown that no physically acceptable states exist with $ E < 0 $. A detailed analysis is presented in Appendix~\ref{app3}, in agreement with the results of Ref.~\cite{Simon08}. Therefore, only states with $ E \geqslant 0 $ are allowed. In the asymptotic regime $ |x| \to \infty $, the $\delta$-function barriers become increasingly sparse and $ V(x) \to 0 $, so that for any positive energy $ E > 0 $, the general solution to the stationary Schr\"odinger equation
\begin{align}
-\psi''(x) + V(x)\psi(x) = E\psi(x)
\end{align}
resembles a free oscillatory wave, which cannot satisfy the square-integrability condition over $ \mathbb{R} $. Consequently, no $ L^2(\mathbb{R}) $ eigenfunction exists for $ E > 0 $.

It has been rigorously shown~\cite{PMAC96} that if the spacing between consecutive point interactions diverges exponentially or faster, as in our model where
$
x_{n+1} - x_n = x_0\lambda^n(\lambda - 1)
$
with $\lambda>1$, then no nontrivial square-integrable solution can exist for any $ E > 0 $. In other words, the positive-energy spectrum contains no embedded eigenvalues. Furthermore, any wavefunction corresponding to
$E>0$ must describe a scattering state. Detailed derivations are provided in Appendix~\ref{app3b}.

For $ E = 0 $, we will demonstrate that a bound-like solution does exist, which satisfies $ \psi(x) \to 0 $ as $ x \to \infty $, though it fails to be square-integrable over the entire real line; for example, it may diverge near the origin. Such a state belongs to the class of ``zero-energy modes" or ``edge states" known from scattering theory: it is a semi-localized or generalized solution, rather than a true eigenstate. This implies that $ E = 0 $ marks the threshold of the continuous spectrum $ [0, \infty) $, but is not an eigenvalue belonging to the discrete spectrum. Nevertheless, the zero-energy wavefunction exhibits a self-similar structure, which warrants further investigation.

\section{The zero-energy wavefunction}\label{Sec3}

We now turn our attention to the zero-energy wavefunction $\psi_0(x)$, which satisfies $H(x)\psi_0(x) = 0$. As previously noted, the scaling symmetry of the Hamiltonian further implies that $H(x)\psi_0(\lambda^{n} x) = 0$ for any $n \in \mathbb{Z}$.
If the potential term is well-behaved, specifically if $V(x)$ is piecewise continuous, the non-degeneracy theorem states \cite{Laudaubook} that the eigenfunction corresponding to a bound-state energy level is unique up to a multiplicative constant; in other words, the energy level is non-degenerate. However, this theorem does not apply in the present context. This raises a question: what is the precise relationship between $\psi_0(x)$ and its scaled version $\psi_0(\lambda^{n} x)$?

\subsection{The recursive matrix}
We begin by deriving recursive formulas for positive $ n $, and then extend these to $ n < 0 $, ensuring consistency of the solution across all $ n \in \mathbb{Z} $. We concentrate on the region $ x > 0 $, with the solution for $ x < 0 $ readily obtainable through the system's symmetry.
Within each interval $ (x_n, x_{n+1}) $, the Schr\"odinger equation simplifies to $ \psi''_0(x) = 0 $, leading us to assume a linear solution of the form:
\begin{align}
\psi_0(x) = a_n x + b_n, \quad x \in (x_n, x_{n+1}).
\end{align}
The boundary conditions at each point $ x = x_n $ are as follows:

1. \textit{Continuity}:
\begin{align}\label{e4}
\psi(x_n^-) = \psi(x_n^+) \implies a_{n-1} x_n + b_{n-1} = a_n x_n + b_n.
\end{align}

2. \textit{Derivative jump due to the $\delta$-function potential}:
\begin{align}
\psi'(x_n^+) - \psi'(x_n^-) = \frac{mV_0}{x_n} \psi(x_n).
\end{align}
Since $ \psi'(x) = a_n $ in $ (x_n, x_{n+1}) $, and $ \psi'(x) = a_{n-1} $ in $ (x_{n-1}, x_n) $, this becomes:
\begin{align}\label{e10}
a_n - a_{n-1} = \frac{mV_0}{x_n} (a_n x_n + b_n).
\end{align}

These equations allow us to compute $ a_n, b_n $ from $ a_{n-1}$, $b_{n-1} $. To make the recursion explicit, solve for $ a_n $ and $ b_n $ simultaneously
from Eqs.(\ref{e4}) and (\ref{e10}), we get
\begin{align}\label{rceqn1}
a_n&=(1+mV_0)a_{n-1}+\frac{mV_0}{x_n}b_{n-1},\notag\\
b_n&=-mV_0x_n a_{n-1}+(1-mV_0)b_{n-1},
\end{align}
or in the matrix form
\begin{align}\label{rceqn2}
\begin{pmatrix}
a_n\\ b_n
\end{pmatrix}
=
\begin{pmatrix}
1+mV_0 & \frac{mV_0}{x_n}\\-mV_0x_n& 1-mV_0
\end{pmatrix}\begin{pmatrix}
a_{n-1}\\ b_{n-1}
\end{pmatrix}
\end{align}
for $n>0$.
Let
\begin{align}
M_n
=
\begin{pmatrix}
1+mV_0 & \frac{mV_0}{x_n}\\-mV_0x_n& 1-mV_0
\end{pmatrix}=\begin{pmatrix}
1+mV_0 & \frac{mV_0}{x_0\lambda^n}\\-mV_0x_0\lambda^n& 1-mV_0
\end{pmatrix}.\notag
\end{align}
It can be verified that $\det M_n = 1$ and $\text{Tr}M_n = 2$, which implies that $M_n$ has a doubly degenerate eigenvalue $\lambda_n = 1$ and is therefore not diagonalizable. Moreover,
\begin{align}
M_n^{-1}
=
\begin{pmatrix}
1-mV_0 & -\frac{mV_0}{x_0\lambda^n}\\mV_0x_0\lambda^n& 1+mV_0
\end{pmatrix}.
\end{align}
(In fact, we have $M_n^{-1}(V_0)=M_n(-V_0)$)
Thus, for $n\leqslant 0$, the recursion relation is
\begin{align}\label{M-1}
\begin{pmatrix}
a_{n-1}\\ b_{n-1}
\end{pmatrix}
=
\begin{pmatrix}
1-mV_0 & -\frac{mV_0}{x_0\lambda^n}\\mV_0x_0\lambda^n& 1+mV_0
\end{pmatrix}\begin{pmatrix}
a_{n}\\ b_{n}
\end{pmatrix}.
\end{align}

\subsection{Bound-like states for $mV_0=1.0$}

We aim to construct a bound-like wavefunction $ \psi_0(x) $ that satisfies $ \psi_0(x) \to 0 $ as $ x \to \infty $.
As previously mentioned, regardless of the value of $ V_0 $, the matrix $ M_n $ has a double eigenvalue equal to 1. This suggests that, in the recursive process, the coefficients may not decay naturally, but instead remain constant or grow linearly. However, by carefully choosing specific initial conditions, we can cancel out this growth, causing both $ a_n $ and $ b_n $ to decay exponentially.

\subsubsection{The recurrence relation and its solution}
To gain insight, we begin with the special case $ mV_0 = 1.0 $, which allows for a simplified analysis. In this case, the recurrence relations reduce to a more tractable form:
\begin{align}
a_n &= 2a_{n-1} + \frac{1}{x_0 \lambda^n} b_{n-1}, \notag \\
b_n &= -x_0 \lambda^n a_{n-1},
\end{align}
for $n>0$.
Eliminating $ b_n $, we obtain a second-order recurrence relation for $ a_n $:
\begin{align}\label{e14}
a_n = 2a_{n-1} - \frac{1}{\lambda} a_{n-2}.
\end{align}
The associated characteristic equation is
\begin{align}\label{e15}
r^2 - 2r + \frac{1}{\lambda} = 0  \implies  r_{1,2} = 1 \pm \sqrt{1 - \frac{1}{\lambda}}.
\end{align}
Since $\lambda>1$, the ``decaying root" $r_2= 1 - \sqrt{1 - \frac{1}{\lambda}}$ satisfies $0<r_2<1$. Thus, if the initial coefficients satisfy $ b_0 = -x_0\lambda \left(1 + \sqrt{1 - \frac{1}{\lambda}}\right) a_0$, it can be found that
\begin{align}\label{e16}
a_n &= a_0r^n_2=a_0 \left(1 - \sqrt{1 - \frac{1}{\lambda}}\right)^n,\notag\\
b_n &= -a_0 x_nr^{n-1}_2=-a_0 x_n \left(1 - \sqrt{1 - \frac{1}{\lambda}}\right)^{n-1}.
\end{align}
Note $0<\lambda r_2=\frac{1}{1+\sqrt{1-\frac{1}{\lambda}}}<1$, then both $a_n$ and $b_n$ decay exponentially as $n\to +\infty$. Therefore, on the interval $(x_n, x_{n+1})$ with $n\geqslant 0$, the wavefunction is given by
\begin{align}\label{e17}
\psi_0(x) = a_n x + b_n=a_0r^{n-1}_2(r_2x-x_n).
\end{align}

Similarly, for $ n \leqslant 0 $, the recursion relation is given by
\begin{align}
\begin{pmatrix} a_{n-1} \\ b_{n-1} \end{pmatrix}
= \begin{pmatrix} 0 & -\frac{1}{x_0 \lambda^n} \\ x_0 \lambda^n & 2 \end{pmatrix}
\begin{pmatrix} a_n \\ b_n \end{pmatrix}.
\end{align}
By eliminating $ b_n $, we derive
\begin{align}
a_{n-2} - 2\lambda a_{n-1} + \lambda a_n = 0,
\end{align}
with the corresponding characteristic equation:
\begin{align}
r^2 - 2\lambda r + \lambda = 0
\implies
r_\pm = \lambda \pm \sqrt{\lambda (\lambda - 1)}.
\end{align}
Comparing with Eq.~(\ref{e15}), we find $ r_+ = \frac{1}{r_2} $ and $ r_- = \frac{1}{r_1} $.
Using the initial condition $ b_0 = -a_0 x_0 r_+ $, we obtain
\begin{align}
a_n &= a_0 r_+^{-n} = a_0 r_2^n, \notag \\
b_n &= -a_0 x_n r_+^{1-n} = -a_0 x_n r_2^{n-1},
\end{align}
which exactly coincides with Eqs.~(\ref{e16}).
Thus, regardless of whether $ n \geqslant 0 $ or $ n < 0 $, $\psi_0(x)$ has a unified expression within the interval $ (x_n, x_{n+1}) $: $\psi_0(x) = a_0 r_2^{n-1} (r_2 x - x_n)$.

\subsubsection{The scaling invariance}

\begin{figure}[th]
\centering
\includegraphics[width=3.6in]{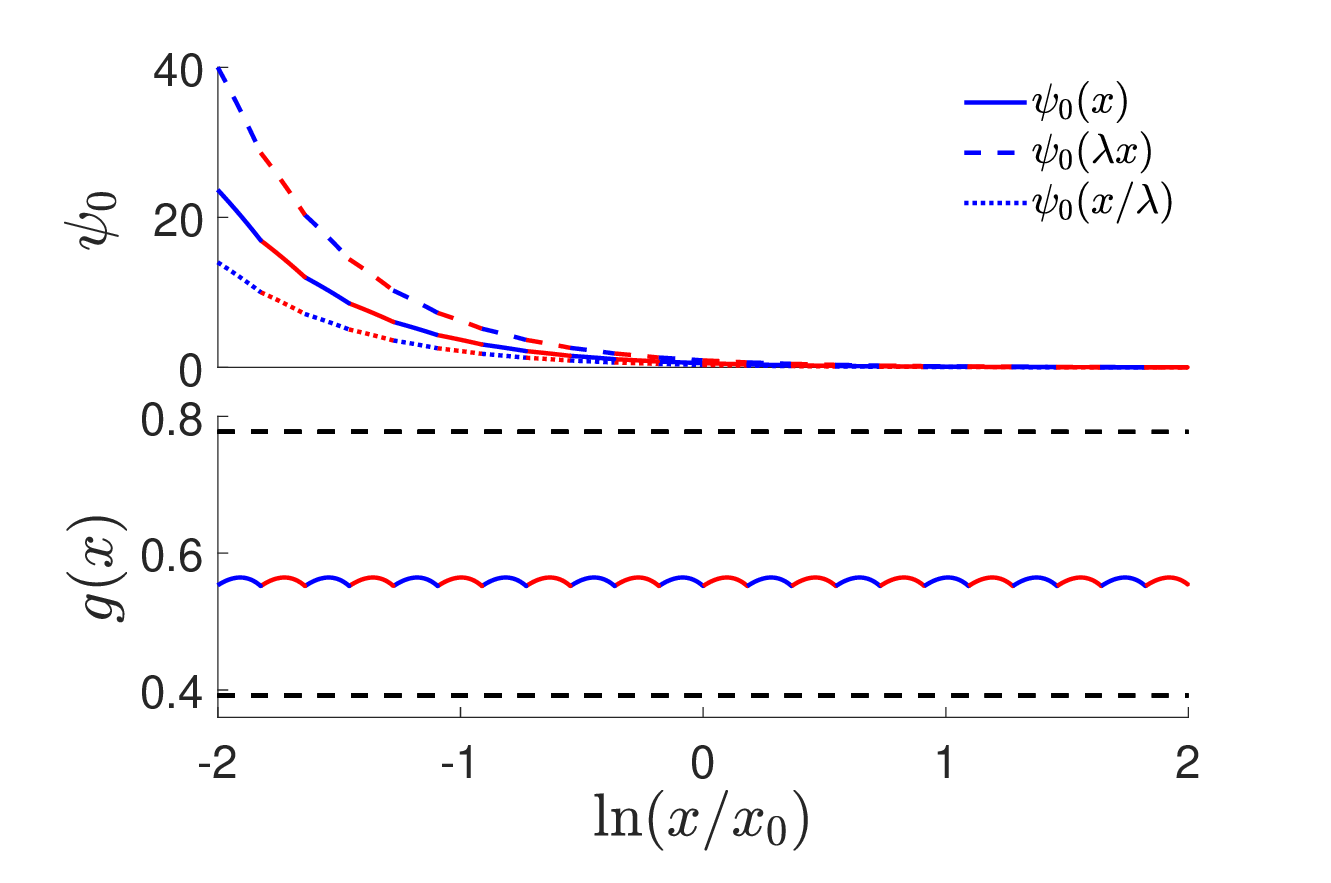}
\caption{
(Upper panel)~$\psi_0(x)$ and its scaled counterparts, $\psi_0(\lambda x)$ and $\psi_0\left(\frac{x}{\lambda}\right)$ versus $\ln(x/x_{0})$;
(Lower panel)~The scaled ratio of the wavefunction $g(x)$ versus $\ln(x/x_{0})$(where we set $\lambda=1.2$, $a_0=-0.8$, $x_0=1$). The upper and lower dashed lines represent $C'\lambda^{\alpha}\approx 0.78$ and $C'\lambda^{-\alpha}\approx 0.39$, respectively.}
\label{Fig1}
\end{figure}

Consider an arbitrary interval $ (x_n, x_{n+1}) $. Under the scaling transformation $ x \to \lambda x $, this interval is mapped to $ (x_{n+1}, x_{n+2}) $, and the corresponding wavefunction becomes:
\begin{align}
\psi_0(\lambda x)& = a_0 r_2^n \left(r_2 \lambda x - x_{n+1} \right)\notag\\&=a_0 r_2^{n} \lambda \left(r_2 x - x_n\right).
\end{align}
Comparing this with the original form of the wavefunction, $ \psi_0(x) = a_0 r_2^{n-1} \left(r_2 x - x_n \right) $ within $ (x_n, x_{n+1}) $, we find
\begin{align}\label{alpha}
\psi_0(\lambda x) = \lambda r_2  \psi_0(x) = \frac{1}{1 + \sqrt{1 - \frac{1}{\lambda}}}  \psi_0(x).
\end{align}
Defining the scaling exponent $\alpha=-\frac{\ln(\lambda r_2)}{\ln\lambda}=\frac{\ln r_1}{\ln\lambda}$, we can express Eq.(\ref{alpha}) in the standard scaling form:
\begin{align}\label{alpha1}
\psi_0(\lambda x) =\lambda^{-\alpha}\psi_0(x).
\end{align}
This demonstrates the discrete scale invariance of the system: the zero-energy wavefunction is self-similar under geometric rescaling of the potential barriers, with its amplitude reduced by a fixed scaling factor. Interestingly, although each segment of $\psi_0(x)$ is piecewise linear with a naive scaling exponent of $1$ (i.e., $f(\lambda x) = \lambda f(x)$), the assembled wavefunction exhibits an overall scaling exponent $-\alpha$. This reflects the influence of the recursive structure of the potential on the global scaling behavior of $\psi_0(x)$. In the upper panel of the Figure~\ref{Fig1}, we plot $\psi_0(x)$, $\psi_0(\lambda x)$, and $\psi_0\left(\frac{x}{\lambda}\right)$ versus $\ln(x/x_{0})$, using distinct colors to highlight their behaviors. Choosing the horizontal axis as $\ln(x/x_0)$ accentuates the self-similar patterns and scaling relationships across different intervals.

\subsubsection{The asymptotic behavior}
According to Eq.(\ref{e17}), at the point $x_n $, the wavefunction is
\begin{align}
\psi_0(x_n) = a_0 \lambda^n r_2^{n-1} x_0(r_2 -1 ) =C (\lambda r_2)^n,
\end{align}
where $C=a_0 x_0(1-\frac{1}{r_2})$. From $x_n=x_0\lambda^n$ we obtain $n =\frac{\ln(x_n/x_0)}{\ln\lambda}$, thereby
\begin{align}
\psi_0(x_n)
= C \me^{n\ln(\lambda r_2)}
= C \Bigl(\frac{x_n}{x_0}\Bigr)^{\frac{\ln(\lambda r_2)}{\ln\lambda}}
=C'x_n^{-\alpha},
\end{align}
where $C'=a_0(1-\frac{1}{r_2})/x_0^{-\alpha-1}$. Hence, it can be shown that, up to an $\mathcal{O}(1)$ factor (for details see Appendix \ref{app4}), the wavefunction exhibits power-law behavior:
\begin{align}\label{psx}
\psi_0(x)\sim x^{-\alpha}.
\end{align}
This exactly agrees with the scaling behavior of  $\psi_0(x)$ in Eq.(\ref{alpha1}).
We define the scaled ratio of the wavefunction as:
\begin{align}\label{D7}
g(x) := \frac{\psi_0(x)}{x^{-\alpha}},
\end{align}
This function encodes the deviation from pure power-law behavior and, due to the discrete scale invariance of the system, satisfies the functional relation $g(\lambda x)=g(x)$, implying log-periodicity. Introducing the logarithmic variable $v =\ln( x / x_0)$, we define $f(\ln(x/x_0))\equiv g(x)$, so that
\begin{align}\label{D7b}
f(v + \ln \lambda) = f(v),
\end{align}
indicating that $f(u)$ is periodic with period $\ln \lambda$. Consequently, $g(x)$ exhibits log-periodic oscillations in $\ln(x/x_0)$, bounded between $C'\lambda^{-\alpha}$ and $C'\lambda^{\alpha}$ (for details see Appendix \ref{app4}). These modulations originate from the self-similar arrangement of the delta-function barriers, where the scaled variable $v = \ln(x / x_0)$ evolves periodically across geometrically spaced intervals. The scaling exponent $\alpha$ captures the global envelope of the wavefunction decay, while its piecewise linear structure leads to multiplicative $\mathcal{O}(1)$ corrections in the form of log-periodic oscillations around the leading $x^{-\alpha}$ behavior. The lower panel of the the Figure~\ref{Fig1} shows $g(x)$ versus the logarithmic coordinate $\ln(x/x_0)$. The curve exhibits clearly visible oscillations that are periodic in $\ln \lambda$, a direct signature of the system’s discrete scale invariance.

\subsubsection{The normalization}

Within $(x_n,x_{n+1})$, the expression of $\psi_0(x)$ can be reexpressed as
\begin{align}\label{psis}
\psi_0(x)=a_0r^{n-1}_2\lambda^n(\lambda^{-n}r_2x-x_0).
\end{align}
To address the normalization of $\psi_0(x)$, we define, for the
$n$-th interval $(x_n,x_{n+1})$:
\begin{align}
x = x_0 \lambda^n u, \quad u \in [1, \lambda], \quad \mathrm{d}x = x_0 \lambda^n \mathrm{d}u.
\end{align}
Using Eq.(\ref{psis}), we obtain:
\begin{align}
&\int_{x_0 \lambda^n}^{x_0 \lambda^{n+1}} \left( a_n x + b_n \right)^2 \mathrm{d}x\notag\\
=& \int_{1}^{\lambda} \left[a_0r^{n-1}_2\lambda^n \bigl( r_2x_0u-x_0 \bigr) \right]^2 \cdot x_0 \lambda^n \mathrm{d}u \notag\\
=&\frac{a^2_0x_0^3}{r^2_2}(\lambda^3r^2_2)^nI
\end{align}
where $I\equiv\mathlarger{\int}_{1}^{\lambda}(r_2u-1)^2\dif u$. Thus,
\begin{align}
\int_{0}^{\infty}|\psi_0(x)|^2\mathrm{d}x
=\frac{a^2_0x_0^3}{r^2_2}I\sum_{n=-\infty}^{+\infty}(\lambda^3r^2_2)^n ,
\end{align}
which is always divergent as long as $\lambda^3r^2_2\neq 0$. Therefore, $ \psi_0(x) $ is not square-integrable. However, since it satisfies $ \psi_0(x) \to 0 $ as $ x \to +\infty $, it can be regarded as a bound-like state.

\subsubsection{Zero-energy wavefunction in momentum space}
Strictly speaking, the full wavefunction $\psi_0(x)$ is non-normalizable and leads to a divergent Fourier transform:
\begin{align}
\tilde{\psi}_0(p) = \frac{1}{\sqrt{2\pi}} \int_{-\infty}^{\infty} \psi_0(x) \me^{-\mi px} \dif x,
\end{align}
as the contribution from the vicinity of $x = 0$ and the accumulation of $\delta$-barriers near the origin cause divergence in both $L^2$ norm and Fourier amplitude.
However, physical observables often probe only a finite or asymptotic spatial region. In particular, the momentum distribution relevant to time-of-flight or Bragg spectroscopy measurements reflects the long-distance behavior of the wavefunction \cite{Bloch2008}. Therefore, we consider the \emph{asymptotic} (cutoff) Fourier component:
\begin{align}\label{F1}
\tilde{\psi}_0^{(>)}(p) = \frac{1}{\sqrt{2\pi}} \int_{x_c}^{\infty} \psi_0(x) \me^{-\mi px} \dif x,
\end{align}
where $x_c$ is a cutoff beyond which $\psi_0(x)$ follows its universal large-$x$ scaling. This cutoff Fourier transform is well-defined and captures the observable infrared behavior of the bound-like state.

To obtain analytical insight, according to Eqs.(\ref{psx}), and (\ref{D7b}) the wavefunction can be expressed as $\psi_0(x) = x^{-\alpha}f(\ln(x/x_0))$. The Fourier transform then becomes
\begin{align}
\tilde{\psi}_0^{(>)}(p) &= \frac{1}{\sqrt{2\pi}} \int_{x_c}^{\infty} x^{-\alpha}f(\ln(x/x_0)) \me^{-\mi px} \dif x.
\end{align}
The asymptotic behavior of this integral depends crucially on the value of the scaling exponent $\alpha$. It can be shown that $\alpha>1$ when $\lambda>\frac{1+\sqrt{5}}{2}$, $\alpha=1$ when $\lambda=\frac{1+\sqrt{5}}{2}$, and $0<\alpha <1$ when $\lambda<\frac{1+\sqrt{5}}{2}$.

\begin{figure}[h]
\centering
\includegraphics[width=3.6in]{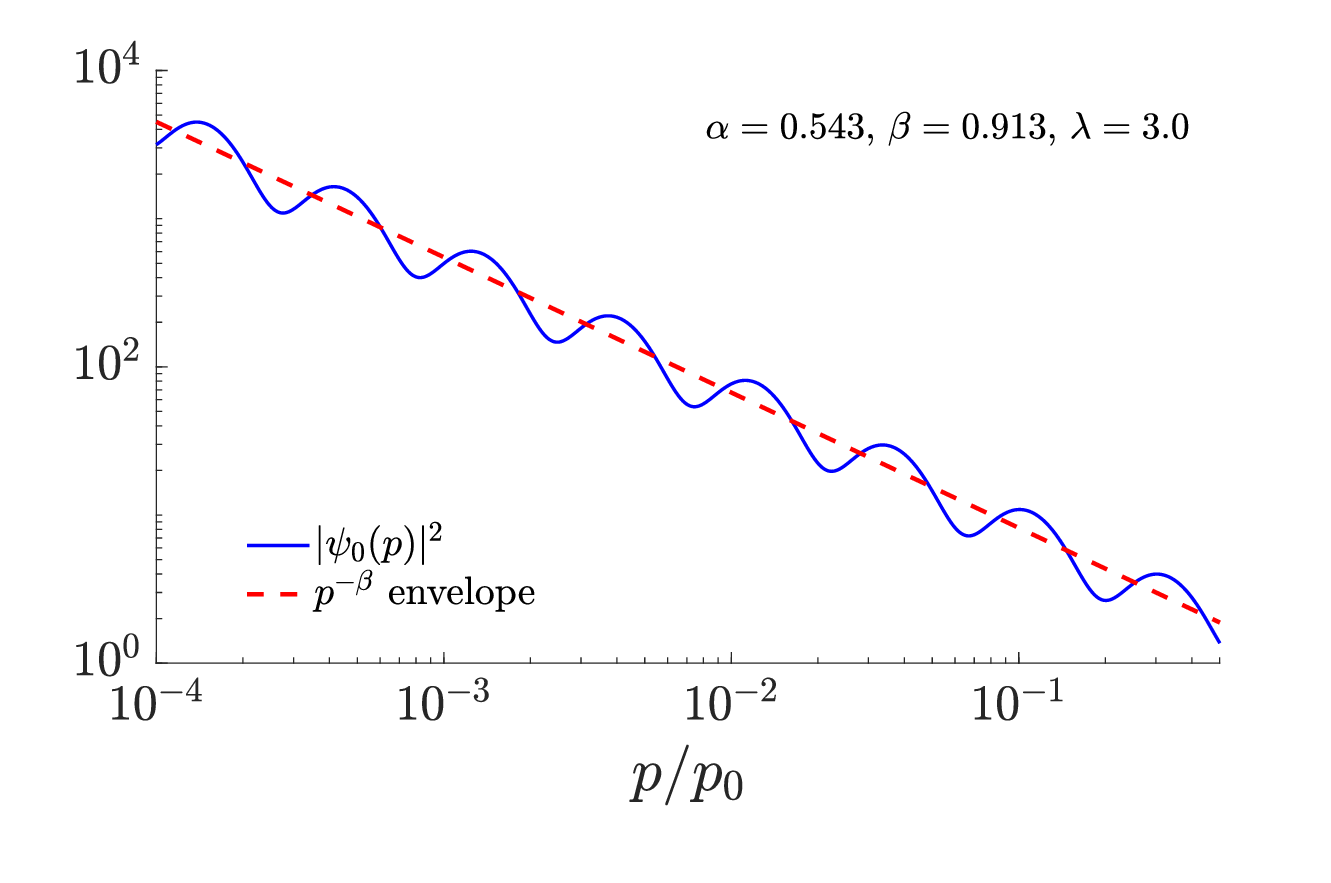}
\caption{
The schematics of momentum distribution in the case of $0<\alpha<1$, where $\lambda=3.0$, $\alpha\approx0.543$, $\beta\approx1.44$. Blue solid curves: the distribution with log-periodic oscillations of period $\ln\lambda$; Red solid curves: theoretical envelopes.
}
\label{Fig2}
\end{figure}
For the case $0 < \alpha < 1$, the Fourier transform exhibits the characteristic threshold singularity:
\begin{align}
&\tilde{\psi}_0^{(>)}(p) = p^{\alpha - 1} F(\ln(p/p_0)), \notag \\
&|\tilde{\psi}^{(>)}_0(p)|^2 = p^{-\beta}F^2(\ln(p/p_0)), \quad \beta = 2(1 - \alpha),
\end{align}
where the function $F(\ln(p/p_0))$, as defined in Eq.(\ref{F}), is a periodic function with period $\ln \lambda$ (for details see Appendix \ref{app4b}). The power-law divergence as $p \to 0$ is a direct consequence of the algebraic decay of the wavefunction in real space. Figure~\ref{Fig2} illustrates the schematics of the momentum distribution $|\tilde{\psi}^{(>)}_0(p)|^2$ in the case of $0<\alpha<1$ where we choose $\lambda=3.0$.  The red curve represents the theoretical envelope capturing the main asymptotic trend, while the blue curves show the distribution with oscillatory modulations that are periodic in $\ln(p/p_0)$, revealing the predicted asymptotic behaviors and log-periodic oscillations.

The analytic structure of the bould-like state in the complex momentum plane further elucidates its properties. The asymptotic form $\tilde{\psi}^{(>)}_0(p) \sim p^{\alpha-1}$ suggests a branch point singularity at $p=0$ when $\alpha-1 \notin \mathbb{Z}$. To confirm this, consider the exact representation:
\begin{align}
\tilde{\psi}^{(>)}_0(p) = \frac{1}{\sqrt{2\pi}} \int_{x_c}^\infty \psi_0(x) \me^{-\mi px} \dif x.
\end{align}
For $\operatorname{Im} p < 0$, the integral converges absolutely due to the exponential decay. Consequently, $\tilde{\psi}^{(>)}_0(p)$ is analytic in the lower half-plane. However, as $p$ approaches the real axis from below, the $p^{\alpha-1}$ behavior emerges, indicating a branch point at $p=0$. This differs fundamentally from the isolated pole singularity $1/(p - \mi 0^+)$ characteristic of true bound states.

When $\alpha\geqslant 1$, the function $F(\ln(p/p_0))$ diverges, and therefore an alternative approach is needed. Since $f(\ln(x/x_0))$ is a periodic function with period $\ln \lambda$, it can be expanded in a Fourier series:
\begin{align}
f(\ln(x/x_0)) = \sum_{k=-\infty}^{\infty} c_k  \me^{\mi k \frac{2\pi}{\ln \lambda} \ln(x/x_0)},
\end{align}
where $c_k$ is the Fourier coefficient of the periodic function $f(\ln(x/x_0))$.

In the case of $\alpha = 1$, corresponding to the critical decay rate, the momentum-space wavefunction exhibits a distinct asymptotic behavior as $p x_c \to 0$:
\begin{align}\label{38b}
\tilde{\psi}_0^{(>)}(p) 
\approx & \frac{c_0}{\sqrt{2\pi}} \left[ -\gamma - \ln(p x_c) + \mi \frac{\pi}{2} \right] \notag\\
&+ \text{periodic terms with period } \ln \lambda.
\end{align}
where $\beta_k=k\frac{2\pi}{\ln \lambda}$, $p_0=1/x_0$ (or $\hbar/x_0$ in SI units), $\gamma$ is the Euler-Mascheroni constant (for details see Appendix \ref{app4b}).
This yields a logarithmic divergence in the momentum distribution and the critical behavior marks the boundary between divergent and convergent infrared properties.

For $\alpha > 1$, reflecting enhanced localization in real space, the Fourier transform approaches a constant at small momenta:
\begin{align}\label{40}
\tilde{\psi}_0^{(>)}(p)\overset{p x_c\to 0} \approx & \frac{1}{\sqrt{2\pi}} \sum_{k=-\infty}^{\infty} c_k x_0^{-\mi \beta_k} \frac{x_c^{1 - \alpha  + \mi \beta_k}}{\alpha - 1 - \mi \beta_k} \notag \\
+&\mathcal{O}(p^{1-\alpha+\mi\beta_{k}}).
\end{align}
 Consequently, the momentum distribution approaches a finite value:
\begin{align}
|\tilde{\psi}^{(>)}_0(p)|^2 \approx |C_0|^2 ,
\end{align}
where $C_0=\frac{1}{\sqrt{2\pi}} \sum_{k=-\infty}^{\infty} c_k x_0^{-\mi \beta_k} \frac{x_c^{1 - \alpha  + \mi \beta_k}}{\alpha - 1 - \mi \beta_k}$.
This convergence reflects the stronger decay of the wavefunction in real space, which suppresses the infrared divergence.

Therefore, the wavefunction in momentum space is given by
\begin{alignat}{2}\label{38}
\tilde{\psi}^{(>)}_0(p)  =& p^{\alpha - 1} F(\ln(p/p_0)) && \quad 0 < \alpha < 1,\notag\\
\tilde{\psi}^{(>)}_0(p) \approx & -\frac{c_0}{\sqrt{2\pi}} \ln p + \text{periodic terms} && \quad \alpha = 1,\notag\\
\tilde{\psi}^{(>)}_0(p) \approx & C_0 && \quad \alpha > 1.
\end{alignat}
Detailed derivations are provided in Appendix \ref{app4b}.

In the complex momentum plane, the analytic structure of the bound-like state depends on $\alpha$. For $0 < \alpha < 1$, it corresponds to a branch point at $p = 0$ on the real axis, reflecting the non-integrable infrared divergence. At the critical value $\alpha = 1$, a logarithmic divergence appears. For $\alpha > 1$, the function is analytic at $p = 0$, with a convergent Taylor series expansion. This distinguishes the state from true bound states with poles off the real axis and scattering states with branch cuts, underscoring its critical, semi-localized character.
The transition at $\alpha = 1$ marks the boundary between infrared-divergent and infrared-convergent momentum behavior, corresponding to a localization transition in real space. This type of correspondence between spatial decay and nonanalyticity in momentum space is well understood in Fourier analysis and asymptotic theory \cite{BenderOrszag1999}.

\subsection{Bound-like states for generic situations}

The above discussion can be extended to more general scenarios with $mV_0 \neq 1.0$, and we summarize the main results below.

From the recurrence relation (\ref{rceqn2}), we obtain
\begin{align}\label{e25}
a_{n+2}
- \Bigl(1 + mV_0 + \frac{1 - mV_0}{\lambda} \Bigr) a_{n+1}
+ \frac{1}{\lambda} a_n
= 0,
\end{align}
and
\begin{align}\label{e26}
b_n = \frac{x_{n+1}}{mV_0} \bigl(a_{n+1} - (1 + mV_0)a_n \bigr).
\end{align}
For the derivation details, please refer to Appendix \ref{app5}.
The characteristic equation associated with Eq.~(\ref{e25}) is
\begin{align}\label{e27}
r^2
- \Bigl(1 + mV_0 + \frac{1 - mV_0}{\lambda} \Bigr) r
+ \frac{1}{\lambda}
= 0,
\end{align}
whose solutions are given by
\begin{align}
\tilde{r}_{1,2} = \frac{1 + mV_0 + \frac{1 - mV_0}{\lambda} \pm \sqrt{\Delta}}{2},
\end{align}
where $\Delta = \left(1 + mV_0 + \frac{1 - mV_0}{\lambda}\right)^2 - \frac{4}{\lambda}$.
Since $0 < \tilde{r}_2 < 1$, we expect the coefficient $a_n$ to take the form $a_n = A \tilde{r}_2^n$, which decays exponentially as $n \to +\infty$. Setting $n = 0$ then this gives $A = a_0$. Substituting it into Eq.~(\ref{e26}), we find that for $n \geqslant 0$,
\begin{align}
a_n &= a_0 \tilde{r}_2^n,\notag\\
b_n &= \frac{x_n}{mV_0} a_0 \tilde{r}_2^{n-1} \left[\tilde{r}_2(1 - mV_0) - 1 \right]\notag\\
&= \frac{x_{n+1}}{mV_0} a_0 \tilde{r}_2^n \left[\tilde{r}_2 - (1 + mV_0) \right],
\end{align}
where in the last line we have used the fact that $\tilde{r}_2$ is a root of Eq.~(\ref{e27}).
When $mV_0 = 1.0$, this result reduces to Eq.~(\ref{e16}). Moreover, it is straightforward to verify that this solution also satisfies the recurrence relation (\ref{M-1}) for $n < 0$. Therefore, within the interval $(x_n, x_{n+1})$, the wavefunction takes the form
\begin{align} \psi_0(x) = a_0 \tilde{r}_2^{n-1}(\tilde{r}_2 x - \tilde{x}_n), \end{align}
where $\tilde{x}_n \equiv \frac{x_n}{mV_0} \left[1 - \tilde{r}_2 (1 - mV_0)\right]$. This expression is a direct generalization of Eq.~(\ref{e16}). Consequently, other properties in the case of $mV_0 \neq 1.0$, such as scaling invariance, power-law behavior, and normalization, can be analyzed in a manner similar to those in the case of $mV_0 = 1.0$.

\subsection{Experimental Implications and Comparison with the Efimov Effect}
The zero-energy wavefunction in our one-dimensional quantum system with self-similar delta-function barriers exhibits discrete scale invariance, a property reminiscent of the Efimov effect. Under spatial rescaling by the factor $\lambda$, the wavefunction transforms as $\psi_0(\lambda x) = \lambda^{-\alpha} \psi_0(x)$, where the scaling exponent $\alpha$ governs the self-similar behavior. For $0 < \alpha < 1$, this mirrors the geometric scaling of Efimov bound states, where energy levels follow a hierarchical progression dictated by a universal scaling constant \cite{EFIMOV1970563, nishida2012}. Additionally, the wavefunction's asymptotic power-law decay, $\psi_0(x) \sim x^{-\alpha}$, resembles the threshold states in Efimov physics, which lie at the boundary between bound and unbound regimes and exhibit weak localization without square-integrability.

In momentum space, the algebraic decay of $\psi_0(x)$ manifests as a threshold singularity at $p=0$, producing a momentum distribution $|\tilde{\psi}^{(>)}_0(p)|^2$ that diverges as $p^{-2(1-\alpha)}$ for $0 < \alpha < 1$, diverges logarithmically for $\alpha = 1$, and approaches a finite limit for $\alpha > 1$. This singularity reflects the extended spatial coherence of the bound-like state and differs fundamentally from true bound-state peaks due to its scaling and lack of Lorentzian lineshape. For $0<\alpha<1$, the behavior is accompanied by log-periodic modulations, appearing as oscillations in $\ln |\tilde{\psi}^{(>)}_0(p)|^2$ versus $\ln p$ with period $\ln \lambda$, arising from coherent interference among the geometrically spaced delta-function barriers, a hallmark of discrete scale invariance also observed in Efimov states and fractal lattice systems \cite{nishida2012}. In experimental realizations using time-of-flight imaging \cite{Bloch2008}, this threshold state would appear as a non-Gaussian tail with superimposed log-periodic modulations rather than a sharp momentum-space peak. Such features are accessible even with modest experimental implementations: as few as 10--20 delta-function potential points in optical lattices may effectively replicate the system's self-similar behavior. By engineering this potential, researchers can directly measure the $\alpha$-dependent singularity and log-periodic oscillations, providing unambiguous evidence of scale invariance.

The kinematically enforced symmetry breaking induced by the singular accumulation of delta-function barriers at $x=0$ further enriches the system's physical implications. This phenomenon, driven by geometric and boundary constraints rather than dynamical interactions, prevents the wavefunction from being symmetrically extended across the origin due to its non-normalizability and the breakdown of continuity at $x=0$. It results in non-symmetric, critical zero-energy states excluded from the symmetric Hilbert space, modifies quantum transition selection rules, and enables asymmetric scattering despite the formally symmetric potential. These effects are conceptually analogous to boundary-induced topological states and symmetry selection in singular quantum systems \cite{simon1976, nishida2012}. By leveraging structural constraints, this symmetry breaking opens avenues for engineering quantum states with tailored properties, distinct from conventional spontaneous symmetry breaking.

Despite these similarities, our model differs significantly from the Efimov effect. The zero-energy wavefunction constitutes a single bound-like state rather than an infinite series of bound states, reflecting the one-dimensional, single-particle nature of our system compared to the three-body correlations central to Efimov physics. For $0 < \alpha < 1$, the scaling behavior resembles Efimov states, but for $\alpha \geq 1$, the momentum-space behavior transitions to logarithmic or finite regimes, further distinguishing the system. The absence of a discrete spectrum in our model, with a continuous spectrum starting at $E=0$, contrasts with the discrete geometric series of Efimov energy levels. These distinctions highlight that our system serves as a minimal realization of discrete scale invariance in one dimension, preserving essential symmetry features while offering analytical simplicity compared to higher-dimensional analogs. The experimental accessibility of its momentum-space signatures and the novel implications of kinematically enforced symmetry breaking position this model as a valuable platform for studying scale-invariant quantum systems.

\section{Conclusion}\label{Sec4}

We have investigated a one-dimensional quantum system with a self-similar potential defined by $V(x) = V_0 \sum_{n=-\infty}^{\infty} \delta(x^2 - x_n^2)$, where $x_n = x_0 \lambda^n$, exhibiting discrete scale invariance. The singular potential induces kinematically enforced symmetry breaking at $x=0$, decoupling spatial regions and resulting in non-symmetric zero-energy states that modify quantum transition selection rules and enable asymmetric scattering. The system possesses a continuous energy spectrum starting at zero, with no discrete bound states. A key feature is the zero-energy wavefunction, which, though not square-integrable, decays to zero at infinity and displays self-similarity under discrete scaling transformations. This behavior, reminiscent of Efimov physics, arises from the recursive structure of the potential but produces a single bound-like state rather than a geometric series of bound states. Using recursive relations, we derived the wavefunction's explicit form, revealing exponential decay across barrier-defined intervals and power-law behavior at large distances for both the special case $mV_0=1.0$ and general positive $V_0$. In momentum space, the wavefunction exhibits a threshold singularity at low momenta, with behavior depending on the scaling exponent $\alpha$, which are experimentally accessible via time-of-flight imaging or momentum-resolved spectroscopy in cold atom systems using optical lattices. This work establishes a minimal yet robust model for exploring scale-invariant quantum systems with singular potentials, providing insights into discrete scale symmetry, its experimental signatures, and potential extensions to scattering studies or higher-dimensional systems.

\section*{ACKNOWLEDGEMENTS}
H.G. was supported by the Innovation Program for Quantum Science and Technology (Grant No. 2021ZD0301904) and the National Natural Science Foundation of China (Grant No. 12074064). X.Y.H. was supported by  the National Natural Science Foundation of China (Grant No. 12405008) and the Jiangsu Funding Program for Excellent Postdoctoral Talent (Grant No. 2023ZB611).

Jia-Chen Tang and Xu-Yang Hou contributed equally to this work.

\appendix
\section{The Scaling Symmetry of the Hamiltonian}\label{app1}
Note the potential term can be expressed by Eq.(\ref{V2}). Under the scaling transformation $ x \to \lambda x $, the potential transforms as
\begin{align}
V(\lambda x) = V_0 \sum_{n} \frac{\delta(\lambda x - x_n) + \delta(\lambda x + x_n)}{2|x_n|}.
\end{align}
Using the scaling property of the delta function
\begin{align}
\delta(\lambda x - x_n) = \frac{1}{\lambda} \delta\left(x - \frac{x_n}{\lambda}\right),
\end{align}
and noting that $ x_n = x_0 \lambda^n = x_{n-1} \lambda $, we have:
\begin{align}
\frac{\delta(\lambda x - x_n)}{2|x_n|}& =  \frac{1}{2\lambda x_n} \delta\left(x - x_0 \lambda^{n - 1}\right) \notag\\&= \frac{1}{2\lambda^2 x_{n-1}} \delta\left(x - x_{n - 1}\right).
\end{align}
Thus,
\begin{align}
V(\lambda x) = \frac{1}{\lambda^2} V(x).
\end{align}
Moreover, the kinetic term $T( x) = - \frac{\dif^2}{\dif x^2}$ transforms as $T(\lambda x) = -\frac{1}{\lambda^2} \frac{\dif^2}{\dif x^2}$ under the transformation $ x \to \lambda x $.
Therefore, the Hamiltonian satisfies $H(\lambda x) = \lambda^{-2} H(x)$.

Suppose an eigenstate satisfies: \begin{align} H(x) \psi(x) = E \psi(x) .\end{align} Now, consider a scaling transformation of the coordinate: $x \to x' = \lambda x$, which implies $x = \frac{x'}{\lambda}$. Define the scaled wavefunction as: $ \psi_\lambda(x') := \psi(\frac{x'}{\lambda})$. We aim to determine whether this scaled function is also an eigenstate, thereby revealing the symmetry properties of $\psi$. After the transformation, we have: \begin{align} H(x') \psi_\lambda(x') = H(x') \psi\left(\frac{x'}{\lambda}\right) .\end{align}
Consider the scaling condition for the Hamiltonian:
\begin{align}
H(\lambda x) = \frac{1}{\lambda^2} H(x).
\end{align}
By substituting $ x = \frac{x'}{\lambda} $, we obtain:
\begin{align}
H(x') = \frac{1}{\lambda^2} H\left(\frac{x'}{\lambda}\right).
\end{align}
Applying this to the eigenvalue equation $ H(x') \psi_\lambda(x') $, we find:
\begin{align}
H(x') \psi_\lambda(x')& = \frac{1}{\lambda^2} H\left(\frac{x'}{\lambda}\right) \psi\left(\frac{x'}{\lambda}\right) \notag\\&= \frac{1}{\lambda^2} E \psi\left(\frac{x'}{\lambda}\right) = \frac{E}{\lambda^2} \psi_\lambda(x').
\end{align}
Thus, the scaled Hamiltonian yields an eigenvalue of $ \frac{E}{\lambda^2}$, demonstrating the effect of the scaling transformation on the energy spectrum. This shows that $\psi_\lambda(x')$ is an eigenstate of the Hamiltonian $H(x')$ with eigenvalue $\frac{E}{\lambda^2}$. In other words, if $\psi(x)$ is an eigenstate of $H(x)$ with energy $E$, then $\psi(\frac{x}{\lambda})$ is an eigenstate with energy $\frac{E}{\lambda^2}$ if it satisfies boundary conditions and normalizability.

\section{The non-existence of bound states with $E < 0$ when $V_0 > 0 $}\label{app3}

The condition for the existence of bound or bound-like states with $ E < 0 $ is as follows.
The Schr\"odinger equation on the interval $ (x_n, x_{n+1}) $ reads:
\begin{align}
-\frac{1}{2m} \psi''(x)  = E \psi(x).
\end{align}
Define $\kappa = \sqrt{2m|E|}$, the solution on $ (x_n, x_{n+1}) $ is given by
\begin{align}
\psi(x) = C_n \mathrm{e}^{-\kappa( x-x_n)} + D_n \mathrm{e}^{\kappa (x-x_n)}.
\end{align}
At $ x = x_n $, the wavefunction satisfies both continuity and derivative jump conditions:

1. Continuity condition:
\begin{align}
C_{n-1} \mathrm{e}^{-\kappa d_n} + D_{n-1} \mathrm{e}^{\kappa d_n} = C_n + D_n,
\end{align}
where $d_n = x_n - x_{n-1} = x_0 \lambda^{n-1} (\lambda - 1)$.

2. Derivative jump condition:
\begin{align}
 -C_n + D_n + C_{n-1} \mathrm{e}^{-\kappa d_n} - D_{n-1} \mathrm{e}^{\kappa d_n}  = \frac{2m V_0}{\kappa x_n} \left( C_n + D_n \right).
\end{align}
As $ n \to +\infty $, the barrier spacing $ d_n \propto \lambda^{n-1} \to \infty $, yielding
\begin{align}
\mathrm{e}^{-\kappa d_n} \to 0, \quad \mathrm{e}^{\kappa d_n} \to \infty.
\end{align}
Thus, the asymptotic forms of the two boundary conditions become
\begin{align}\label{bc1}
D_{n-1} \mathrm{e}^{\kappa d_n} \approx C_n + D_n,
\end{align}
and
\begin{align}\label{bc2}
 -C_n + D_n - D_{n-1} \mathrm{e}^{\kappa d_n}  \approx \frac{2m V_0}{\kappa x_n} \left( C_n + D_n \right).
\end{align}
Using Eq.(\ref{bc1}) to eliminate $C_n$, Eq.(\ref{bc2}) yields
\begin{align}
D_n \approx \left( \frac{m V_0}{\kappa x_n} + 1 \right) D_{n-1} \mathrm{e}^{\kappa d_n}\approx D_{n-1} \mathrm{e}^{\kappa d_n}
\end{align}since $\frac{m V_0}{\kappa x_n} \to 0$. Substitute $ d_n = x_0 \lambda^{n-1} (\lambda - 1) $ to get the recurrence:
\begin{align}
D_n \approx D_{n-1} \mathrm{e}^{\kappa x_0 (\lambda - 1) \lambda^{n-1}}.
\end{align}
Since $ \lambda > 1 $, the exponential factor $ \kappa x_0 \lambda^n $ grows exponentially with $ n $, leading to $ D_n \propto \mathrm{e}^{\kappa x_0 \lambda^n} \to \infty $.
Therefore, no bound or bound-like state solution exists. Strictly speaking, there does not even exist a physically acceptable solution, because any solution $ \psi_E(x) $ with $ E < 0 $ satisfies $ |\psi_E(x)| \to \infty $ as $ x \to \infty $.

\section{The wavefunction associated with $E>0$}\label{app3b}

Based on the previous discussion, for $E > 0$, the wavefunction in the interval $(x_n, x_{n+1})$ takes the form
\begin{align} \psi_E(x) = A_n \me^{\mi k(x - x_n)} + B_n \me^{-\mi k(x - x_n)}, \end{align} where $k=\sqrt{2mE}$.
At $x = x_n$, the continuity condition gives \begin{align}  A_{n-1} \me^{\mi k d_{n}} + B_{n-1} \me^{-\mi k d_{n}}=A_n + B_n, \end{align}
and the jump condition for the derivative leads to
\begin{align} A_n - B_n - A_{n-1} \me^{\mi k d_{n}} + B_{n-1} \me^{-\mi k d_{n}} = \frac{ m V_0}{\mi k x_n} (A_n + B_n). \end{align}
Rearranging these equations, we obtain following transfer matrix relation:
\begin{align}
\begin{pmatrix}
A_n \\
B_n
\end{pmatrix}
=
\begin{pmatrix}
\me^{\mi\theta_{n}} (1 + \eta_n) & \eta_n \me^{-\mi\theta_{n}} \\
-\eta_n \me^{\mi\theta_{n}} & \me^{-\mi\theta_{n}} (1 - \eta_n)
\end{pmatrix}
\begin{pmatrix}
A_{n-1} \\
B_{n-1}
\end{pmatrix},
\end{align}
where $\theta_n = k d_n$ and $\eta_n = \frac{m V_0}{2\mi k x_n}$.
As $n \to +\infty$, we have $x_n \to \infty$, and the barrier spacing $d_n = x_0 \lambda^{n-1}(\lambda - 1)$ increases exponentially. Simultaneously, the barrier strength $\frac{m V_0}{x_n}$ tends to zero. In this limit, $\eta_n \to 0$, and the transfer matrix reduces to
\begin{align}
\begin{pmatrix}
A_n \\
B_n
\end{pmatrix}
\approx
\begin{pmatrix}
\me^{\mi\theta_{n}} & 0 \\
0 & \me^{-\mi\theta_{n}}
\end{pmatrix}\begin{pmatrix}
A_{n-1} \\
B_{n-1}
\end{pmatrix},
\end{align}
i.e., $A_n \approx \me^{\mi\theta_n} A_{n-1}$ and $B_n \approx \me^{-\mi\theta_n} B_{n-1}$.
Therefore, the amplitudes $|A_n|$ and $|B_n|$ remain constant and do not decay exponentially. As $n \to +\infty$, the potential becomes increasingly sparse and asymptotically approaches free space. In this regime, the wavefunction behaves as a plane wave corresponding to scattering states in the continuous spectrum, making localization impossible.

\section{Power-law approximation of $\psi_0(x)$}\label{app4}
Within the $n$-th interval $x_n \leqslant x \leqslant x_{n+1} = \lambda x_n$,
the wavefunction is linear, given by $\psi_0(x) = a_n x + b_n$. At the interval endpoints, it takes the values
\begin{align} \psi_0(x_n) &= C' x_n^{-\alpha}, \notag \\ \psi_0(x_{n+1}) &= C' x_{n+1}^{-\alpha} = C' (\lambda x_n)^{-\alpha} = C' \lambda^{-\alpha} x_n^{-\alpha}, \end{align}
where $\alpha = -\frac{\ln(\lambda r_2)}{\ln \lambda} > 0$. Let $a_0<0$, then $\psi_0(x)$ is linear and decreasing within $[x_n, x_{n+1}]$, and its value lies between the endpoint values:
\begin{align} C' \lambda^{-\alpha} x_n^{-\alpha} \leqslant \psi_0(x) \leqslant C' x_n^{-\alpha}. \end{align}
To express $\psi_0(x)$ relative to $x^{-\alpha}$, consider $x \in [x_n, x_{n+1}]$, where $x_n = x_0 \lambda^n$ and $x_{n+1} = \lambda x_n$. Since $\alpha > 0$, the function $x^{-\alpha}$ is decreasing, so:
\begin{align} x_{n+1}^{-\alpha} = \lambda^{-\alpha} x_n^{-\alpha} \leqslant x^{-\alpha} \leqslant x_n^{-\alpha}. \end{align}
Writing $x = x_n u$ with $u \in [1, \lambda]$, we have $x_n = x / u$ and $x_n^{-\alpha} = u^{\alpha} x^{-\alpha}$, so the bounds become:
\begin{align} \psi_0(x) &\leqslant C' x_n^{-\alpha} = C' u^{\alpha} x^{-\alpha}, \notag\\\psi_0(x) &\geqslant C' \lambda^{-\alpha} x_n^{-\alpha} = C' \lambda^{-\alpha} u^{\alpha} x^{-\alpha}. \end{align}
Since $u^{\alpha} \in [1, \lambda^{\alpha}]$, it follows that:
\begin{align} C' \lambda^{-\alpha} x^{-\alpha} \leqslant \psi_0(x) \leqslant C' \lambda^{\alpha} x^{-\alpha}. \end{align}
Thus, the ratio $\frac{\psi_0(x)}{x^{-\alpha}}$ is bounded:
\begin{align}
&\inf \frac{\psi_0(x)}{x^{-\alpha}} \geqslant C'\lambda^{-\alpha} > 0, \notag\\
&\sup \frac{\psi_0(x)}{x^{-\alpha}} \leqslant C'\lambda^{\alpha} < \infty,
\end{align}
Since $C'$, $\lambda$, and $\alpha$ are all independent of $x$, the ratio $\frac{\psi_0(x)}{x^{-\alpha}}$ remains bounded within the fixed constant interval $[C' \lambda^{-\alpha},\; C' \lambda^{\alpha}]$ throughout the entire domain. It neither diverges nor vanishes as $x$ increases. This establishes that the leading-order behavior of $\psi_0(x)$ follows a power-law decay $\psi_0(x) \sim x^{-\alpha}$, up to a multiplicative correction of order $\mathcal{O}(1)$.

\section{Some derivations about the Fourier transformation of $\psi_0(x)$}\label{app4b}
Using $\psi_0(x)=x^{-\alpha}f(\ln(x/x_0))$, the Fourier transformation in Eq.(\ref{F1}) yields
\begin{align}
\tilde{\psi}^{(>)}_0(p) = \frac{1}{\sqrt{2\pi}} \int_{x_c}^\infty x^{-\alpha} f(\ln(x/x_0)) \me^{-\mi px} \dif x.
\end{align}
Change variable $x = y/p$ to extract the scaling structure:
\begin{align}
\tilde{\psi}^{(>)}_0(p)
&= \frac{p^{\alpha - 1}}{\sqrt{2\pi}}  \int_{p x_c}^\infty y^{-\alpha} f(\ln(y) - \ln(p/p_0)) \me^{-\mi y} \dif y,
\end{align}
where $p_0=1/x_0$.
Define the function
\begin{align}\label{F}
F(\ln(p/p_0)) := \frac{1}{\sqrt{2\pi}}\int_{p x_c}^\infty y^{-\alpha} f(\ln y - \ln(p/p_0))  \me^{-\mi y}  \dif y,
\end{align}
which is finite for $0 < \alpha < 1$.
Then, we obtain
\begin{align}
\tilde{\psi}^{(>)}_0(p) = p^{\alpha - 1} F(\ln(p/p_0)).
\end{align}
Since $f(\ln(x/x_0))$ is periodic in $\ln(x/x_0)$ with period $\ln \lambda$, the shifted argument leads to:
\begin{align}
&F\left(\ln\frac{p}{p_0} + \ln \lambda\right) \notag\\
&= \frac{1}{\sqrt{2\pi}}\int_{p x_c}^\infty  f\left(\ln y -\ln\frac{p}{p_0}  - \ln \lambda\right) \frac{\me^{-\mi y}}{y^\alpha} \dif y \notag\\
&= F\left(\ln\frac{p}{p_0} \right).
\end{align}
Hence, $F(\ln(p/p_0))$ is a periodic function with period $\ln \lambda$.

When $\alpha\geqslant 1$, Eq.(\ref{F}) diverges, and therefore an alternative approach is needed. Since $f(\ln(x/x_0))$ is a periodic function with period $\ln \lambda$, it can be expanded in a Fourier series:
\begin{align}
f(\ln(x/x_0)) = \sum_{k=-\infty}^{\infty} c_k  \me^{\mi k \frac{2\pi}{\ln \lambda} \ln(x/x_0)}.
\end{align}

First, we consider the case of $\alpha=1$. Define $\beta_k = k \frac{2\pi}{\ln \lambda}$, and use $\me^{\mi k \frac{2\pi}{\ln \lambda} \ln(x/x_0)} = \left( \frac{x}{x_0} \right)^{\mi \beta_k}$,
then
\begin{align}
\psi_0(x) = x^{-1} \sum_{k=-\infty}^{\infty} c_k \left( \frac{x}{x_0} \right)^{\mi \beta_k}.
\end{align}
Substituting into the Fourier transform, yielding
\begin{align}
\tilde{\psi}_0^{(>)}(p) = \frac{1}{\sqrt{2\pi}} \sum_{k=-\infty}^{\infty} c_k x_0^{-\mi \beta_k} \int_{x_c}^{\infty} x^{-1 + \mi \beta_k} \me^{-\mi p x}  \dif x.
\end{align}
Introducing the integral $I_k(p) = \mathlarger{\int}_{x_c}^{\infty} x^{-1 + \mi \beta_k} \me^{-\mi p x}  \dif x$ and $t = p x$, we have
\begin{align}
I_k(p) =& \int_{p x_c}^{\infty} \left(\frac{t}{p}\right)^{-1 + \mi \beta_k} \me^{-\mi t} \frac{\dif t}{p}\notag\\
=& p^{-\mi \beta_k} \int_{p x_c}^{\infty} t^{-1 + \mi \beta_k} \me^{-\mi t}  \dif t.
\end{align}
We focus on the asymptotic behavior of the integral in the $p x_c \to 0^+$ limit. For $k = 0$ (i.e., $\beta_0 = 0$):
\begin{align}
&\int_{p x_c}^{\infty} t^{-1} \me^{-\mi t}  \dif t = - \text{Ei}(-\mi p x_c) \notag \\
=& - \left[ \gamma + \ln(p x_c) - \mi \frac{\pi}{2} \right] + \mathcal{O}(p),
\end{align}
where $\gamma$ is the Euler-Mascheroni constant and $\text{Ei}(z)=-\mathlarger{\int}_{-z}^{\infty}\dfrac{\me^{-t}}{t}\dif t$ is the exponential integral function. For $k \neq 0$, the integral converges and introduces a periodic term:
\begin{align}\label{Ik}
\tilde{I}_k = \int_{0}^{\infty} t^{-1 + \mi \beta_k} \me^{-\mi t}  \dif t + \mathcal{O}(p),
\end{align}
which is well-defined due to the oscillatory nature of the integrand. In total:
\begin{align}
\tilde{\psi}_0^{(>)}(p) \approx & \frac{c_0}{\sqrt{2\pi}} \left[ -\gamma - \ln(p x_c) + \mi \frac{\pi}{2} \right] \notag\\
+& \frac{1}{\sqrt{2\pi}} \sum_{k \neq 0} c_k x_0^{-\mi \beta_k} p^{-\mi \beta_k} \tilde{I}_k.
\end{align}
Moreover, $p^{-\mi \beta_k} = \me^{-\mi \beta_k \ln p}$, which has periodicity $\ln p + \frac{2\pi}{\beta_k} = \ln p + \ln \lambda$. Therefore, when $\alpha = 1$,
\begin{align}
\tilde{\psi}_0^{(>)}(p) \approx c_0 \ln p + \text{periodic terms with period } \ln \lambda.
\end{align}

When $\alpha > 1$, we have
\begin{align}
\psi_0(x) = x^{-\alpha} \sum_{k=-\infty}^{\infty} c_k x_0^{-\mi \beta_k} x^{\mi \beta_k},
\end{align}
and therefore
\begin{align}
\tilde{\psi}_0^{(>)}(p) = \frac{1}{\sqrt{2\pi}} \sum_{k=-\infty}^{\infty} c_k x_0^{-\mi \beta_k} \int_{x_c}^{\infty} x^{-\alpha + \mi \beta_k} \me^{-\mi p x} \dif x.
\end{align}
The integral can be expressed in terms of the upper incomplete Gamma function
\begin{align}
&\int_{x_c}^{\infty} x^{-\alpha + \mi \beta_k} \me^{-\mi p x} \dif x \notag \\
&=(\mi p)^{\alpha-\mi \beta_{k}-1}\Gamma(1-\alpha+\mi\beta_{k},\mi p x_{c}),
\end{align}
in the limit $p x_{c} \to 0$, $\Gamma(1-\alpha+\mi\beta_{k},\mi p x_{c})$ reduces to
\begin{align}
&\Gamma(1-\alpha+\mi\beta_{k},\mi p x_{c}) \approx -\frac{(\mi p x_{c})^{1-\alpha+\mi\beta_{k}}}{1-\alpha+\mi\beta_{k}} \notag \\
&+\Gamma(1-\alpha+\mi\beta_{k})+\mathcal{O}((\mi p x_{c})^{2-\alpha+\mi\beta_{k}})
\end{align}
and thereby
\begin{align}
\tilde{\psi}_0^{(>)}(p)\overset{p x_c\to 0}  \approx & \frac{1}{\sqrt{2\pi}} \sum_{k=-\infty}^{\infty} c_k x_0^{-\mi \beta_k} \frac{x_c^{1 - \alpha + \mi \beta_k}}{\alpha - 1 - \mi \beta_k} \notag \\
+&\mathcal{O}(p^{1-\alpha+\mi\beta_{k}}),
\end{align}
which is approximately a constant, in agreement with Eq.(\ref{40}).


\section{Generic recurrence relation}\label{app5}
From Eq.(\ref{rceqn2}), we obtain
\begin{align}
b_{n-1}
= \frac{x_n}{mV_0} \Bigl(a_n - (1 + mV_0)a_{n-1} \Bigr)\,.
\end{align}
Substituting this expression into
\begin{align}
b_n = -mV_0\,x_n\,a_{n-1} + (1 - mV_0)\,b_{n-1},
\end{align}
yields
\begin{align}
b_n
&= -mV_0\,x_n\,a_{n-1}\notag\\
&+ (1 - mV_0)\frac{x_n}{mV_0} \Bigl[a_n - (1 + mV_0)a_{n-1} \Bigr].
\end{align}
On the other hand, from the recurrence at the next step, we have
\begin{align}
&a_{n+1} = (1 + mV_0)a_n + \frac{mV_0}{x_{n+1}}b_n\notag\\
\implies
&b_n = \frac{x_{n+1}}{mV_0} \bigl(a_{n+1} - (1 + mV_0)a_n \bigr).
\end{align}
Equating the two expressions for $ b_n $, and using the relation $ x_{n+1}/x_n = \lambda $, we finally get
\begin{align}
a_{n+2}
- \Bigl(1 + mV_0 + \frac{1 - mV_0}{\lambda} \Bigr) a_{n+1}
+ \frac{1}{\lambda} a_n
= 0.
\end{align}
When $ mV_0 = 1.0 $, it reduces to Eq.(\ref{e14}).

\bibliographystyle{apsrev}
\bibliography{Review1,Review2}

\end{document}